\documentclass[english]{emulateapj}

\usepackage{textcomp}
\usepackage{amsmath,amssymb,amsfonts}
\usepackage{hyperref}
\usepackage{txfonts}
\usepackage{subfigure}
\usepackage{natbib}

\newcommand{\di}{\partial} 
\def\Hcal{{\cal H}}

\def\Hcal{{\cal H}}  
 \def\Acal{{\cal A}}

\begin{document}

 \title{\bf{Post main sequence evolution of icy minor planets: implications for water retention and white dwarf pollution}}

\author{Uri Malamud and Hagai B. Perets}
\affil{Department of Physics, Technion, Israel}
\email{uri.mal@tx.technion.ac.il ~~~~ hperets@physics.technion.ac.il}

\newpage

\begin{abstract}
Most observations of polluted white dwarf atmospheres are consistent with accretion of water depleted planetary material. Among tens of known cases, merely two cases involve accretion of objects that contain a considerable mass fraction of water. The purpose of this study is to investigate the relative scarcity of these detections. Based on a new and highly detailed model, we evaluate the retention of water inside icy minor planets during the high luminosity stellar evolution that follows the main sequence. Our model fully considers the thermal, physical, and chemical evolution of icy bodies, following their internal differentiation as well as water depletion, from the moment of their birth and through all stellar evolution phases preceding the formation of the white dwarf. We also account for different initial compositions and formation times. Our results differ from previous studies, that have either underestimated or overestimated water retention. We show that water can survive in a variety of circumstances and in great quantities, and therefore other possibilities are discussed in order to explain the infrequency of water detections. We predict that the sequence of accretion is such that water accretes earlier, and more rapidly than the rest of the silicate disk, considerably reducing the chance of its detection in H-dominated atmospheres. In He-dominated atmospheres, the scarcity of water detections could be observationally biased. It implies that the accreted material is typically intrinsically dry, which may be the result of inside-out depopulation sequence of minor planets. 
\end{abstract}


\section{Introduction}\label{S:Intro}
The sinking time scale of elements heavier than helium in the atmospheres of WDs are relatively short \citep{Koester-2009}, ranging from several days in H-dominated WDs to no more than $\sim$1 Myr in He-dominated WDs. Nevertheless, between 25\% to 50\% of all WDs \citep{ZuckermanEtAl-2003,ZuckermanEtAl-2010,KoesterEtAl-2014} are polluted with heavy elements. Since WDs with a cooling age of more than a few tens of Myr are not sufficiently hot to radiatively levitate heavy elements from the core, this observation is suggestive of their ongoing accretion. Based on several lines of arguments \citep{FarihiEtAl-2010,KoesterEtAl-2014,VerasGansicke-2015} it is now clear that this accretion is not from the interstellar medium, and the prevailing paradigm is that it is indicative of accretion of planetary material \citep{DebesSigurdsson-2002,Jura-2003,KilicEtAl-2006,Jura-2008}. Although only a small percentage of WDs show infrared excess consistent with a disk \citep{BarberEtAl-2012,Farihi-2016}, in the last decade, nearly 40 dusty disks have nevertheless been detected around polluted WDs \citep{RocchettoEtAl-2015,Farihi-2016}, and less frequently disks containing gas \citep{WilsonEtAl-2014,ManserEtAl-2016}. In contrast, there is no confirmed detection of a single debris disk around an unpolluted WD \citep{XuEtAl-2015}.

The current picture is that some minor planets survive the RGB and AGB stellar evolution phases and are subsequently perturbed to orbits with a pericentre inside the tidal disruption radius of the WD \citep{DebesSigurdsson-2002,BonsorEtAl-2011,DebesEtAl-2012,FrewenHansen-2014,MustillEtAl-2014,VerasGansicke-2015}, turn into a circumstellar disk \citep{VerasEtAl-2014b,VerasEtAl-2015}, and eventually accrete onto the WD by various mechanisms \citep{Jura-2008,Rafikov-2011,MetzgerEtAl-2012}. Recently,  \cite{VanderburgEtAl-2015} have identified a cloud of debris in the process of disintegrating \citep{GaensickeEtAl-2015,XuEtAl-2016}. 

An alternative option is instantaneous accretion via a direct impact onto the WD, as opposed to accretion from a disk, which persists for $10^4$-$10^6$ years \citep{GirvenEtAl-2012}. According to \cite{WyattEtAl-2014} and \cite{VerasEtAl-2014b}, direct accretion is far less likely, since the WD radius is about $\sim$70 times smaller than the tidal disruption radius.

Spectroscopic analysis of WD atmospheres \citep{WolffEtAl-2002,DufourEtAl-2007,DesharnaisEtAl-2008,KleinEtAl-2010,GansickeEtAl-2012} as well as infrared spectroscopy of the debris disks themselves \citep{ReachEtAl-2005,JuraEtAl-2007,ReachEtAl-2009,JuraEtAl-2009,BergforsEtAl-2014}, are overall consistent with dry compositions, characteristic of inner solar system objects. In particular, WD atmospheres are notably depleted of volatile elements. Only two polluted He-dominated WDs are potentially water rich (abundant refractory elements and trace oxygen in excess of that expected for metal oxides only). Known cases are limited to 26\% water in GD 61 \citep{FarihiEtAl-2013} and 38\% water in SDSSJ1242 \citep{RaddiEtAl-2015}. Also, by looking at en ensemble of 57 nearby He-dominated WDs, \cite{JuraXu-2012} concluded that the summed hydrogen in their atmospheres must have been delivered by very dry bodies, the water mass fraction not exceeding 1\%. However, only a handful of these nearby WDs have actually been examined with high spectral sensitivity and reported to have oxygen. By looking at this particular oxygen-containing subset of WDs, \cite{JuraYoung-2014} further show that no more than 5.8\% of this oxygen could have been carried in water. A common bias of these studies is that they involve relatively young WDs with cooling ages up to $\sim$300 Myr \citep{VerasEtAl-2014c}. 

A different line of investigation may support the view that in the long term, WDs are being polluted by a continuous flux of water bearing bodies \citep{RaddiEtAl-2015}. The evidence for this comes from He-dominated WDs with trace hydrogen, that observationally increases with cooling age \citep{DufourEtAl-2007}. Although biased by both physical effects (the depth of the convection zone) and observational limitations, according to \cite{VerasEtAl-2014c} it could imply the long term accretion of water-bearing planetesimals, and it is not likely to be attributed to primordial hydrogen or interstellar accretion \citep{BergeronEtAl-2011}. Polluted He-dominated (DBZ) WDs like GD 61 and SDSSJ1242 will eventually resemble typical DBA WDs (with some hydrogen but no heavy elements), as their heavy elements will diffuse out of the convection zone.

We might conclude that accretion events of water-bearing bodies onto WDs are compatible with the long term trace hydrogen trend, however that these events are relatively rare, and are not frequently observed on an individual basis. The purpose of this study is to investigate why. First, we reevaluate, based on a new and improved model, how much water can survive the high luminosity RGB and AGB stellar evolutionary phases. Early work on this problem includes the seminal studies of \cite{SternEtAl-1990} and \cite{JuraXu-2010}, however the models they use have various limitations that need to be addressed. Based on our findings, we then discuss different potential explanations for the scarcity of water detections. The paper is arranged as follows. In Section \ref{S:Models} we outline the primary aspects of the two models used in previous studies and compare them to our model, demonstrating the various improvements and study assumptions. In Section \ref{S:Results} the results of our model are presented, and discussed in Section \ref{S:discussion}. We conclude with a summary of our findings in Section \ref{S:Conclusions}. 

\section{Computer models}\label{S:Models}
\subsection{Previous studies}\label{SS:Previous}
The study by \cite{SternEtAl-1990} dates back to the era before exo-planet discoveries. The authors suggest that some volatile emissions commonly observed around post main sequence stars can be explained by sublimation of large reservoirs of comets during the high luminosity stellar evolution phases following the main sequence, hence indirectly supporting the existence of exo-comets, and by extension, planetary systems. They consider primarily comet sized objects, although larger sizes are also discussed. In order to calculate the sublimation rate, they use a simple model. The rate is obtained as a function of the global average surface temperature, prompted by external insolation, and accounting for the effect of latent heat released by sublimation. The object is assumed to lose mass as its surface is sublimating away and its size diminishing, until finally reducing to zero. The authors do not consider the formation of a lag-deposit (mantling effect) or internal heat transport. The actual evolution is calculated only for the horizontal-branch phase for a period of $\sim 10^7$ yr, and the luminosity is simply assumed to be constant during that period. The result of the model is complete sublimation of all comets within approximately 300 AU of the parent star, which is a considerable over-estimation, as also indicated by more recent exo-comet sublimation studies \citep{Jura-2004}.

The first model to consider the post main sequence retention of water in the interior of exo-solar icy bodies has been developed by \cite{JuraXu-2010}. The authors recognized that if a planetary object is sufficiently large, then the stellar evolution after the main sequence, however luminous, may not be long enough for the heat to penetrate and clear out all the internal water. Their model assumes that heat is transported to the interior by conduction, and that if at a given location the temperature increases to the point that water becomes gaseous it is immediately and completely expelled from the body. The internal evolution effectively starts after the main sequence, since radiogenic heating is not considered, nor is any other internal heat source. The body is assumed to initially have a uniform composition and temperature. The model also assumes a constant thermal diffusivity coefficient taken to represent typical terrestrial rocks. The size range investigated is above that of comets, the radius ranging from 10 km to 200 km. The density is assumed to be equal to that of Ceres (2.1 gr cm$^{-3}$). Unlike in earlier studies, \cite{JuraXu-2010} consider the exact time variable luminosity obtained from stellar evolution models (the solar masses investigated are 1 M$_{\odot}$ and 3 M$_{\odot}$). They also account for the mass loss of the parent star during the AGB phase and, neglecting drag, consistently calculate the change in the orbit of the icy object around the star as it conserves its angular momentum. Their results show that in this investigated size range icy objects are large enough to retain much of their initial water, even with a semi major axis as close as 5 AU. 

Although the model of \cite{JuraXu-2010} represents a major step forward compared to previous studies, it still has several shortcomings. One of the primary concerns is the effect that the evolution during the main sequence has on the evolution after the main sequence. While a uniform internal temperature is a reasonable initial condition for small and intermediate sized objects after several Gyr of thermal evolution, a uniform, unaltered composition is not. In fact, in objects larger than comets, water/rock internal differentiation during the main sequence could result in significant changes to both internal composition and structure \citep{MalamudPrialnik-2013}. This heterogeneity can greatly influence the final fraction of water surviving in the WD's planetary system. Below we list a number of important parameters that need to be considered:

(a) \emph{Initial composition} - if the body heats sufficiently during the main sequence and differentiates to form a stratified structure, it is easier to remove the water during the RGB and ABG phases, since it is much closer to the surface and thus the heat has a shorter distance to penetrate. A higher initial rock/ice mass ratio would naturally lead to the formation of a thinner ice mantle, and it would also increase the amount of internal radiogenic heating. \cite{JuraXu-2010} assume that the rock/ice mass ratio is 1, however in solar system objects the ratio for outer solar system and Kuiper belt objects (KBOs) could also be much higher (see Section \ref{SS:Configuration}). 

(b) \emph{Formation time} - radiogenic heating is in itself related to the formation time of an object. Objects that form as far as the Kuiper belt have formation times of at least several tens of Myr \citep{KenyonBromley-2012} and are therefore only affected by long-lived radionuclides. At a distance of only 5-10 AU from the star however, evidence suggests that the formation time could be of the order of only 2-5 Myr after the formation of calcium aluminum inclusions (CAIs). This is inferred from observations of protoplanetary disks \citep{HartmannEtAl-1998,HaischEtAl-2001}, as well as from chondritic meteorites \citep{KleineEtAl-2005} and from Mars meteorites \citep{DauphasPourmand-2011}, since Mars is believed to be a standard planetary embryo and probably formed on the same time scale. Objects that form at these distances may thus be subject to very intense short-lived radiogenic heating during the first few Myr of their evolution, even if they are otherwise too small to be affected by long-term radiogenic heating. 

(c) \emph{Chemical reactions} - in addition to radiogenic heat, the heat released by chemical reactions via the interaction between liquid water and rock (known as serpentinization reactions) is not negligible \citep{JewittEtAl-2007}. The even more significant contribution of serpentinization is that it alters the rock on the molecular level, embedding water onto its structure \citep{MalamudPrialnik-2013}. In recent years modeling serpentinization reactions inside icy bodies has attracted a lot of attention \citep{MalamudPrialnik-2013,HolmEtAl-2015,MalamudPrialnik-2015,NeveuEtAl-2015,TravisSchubert-2015,MalamudPrialnik-2016},  and in the specific case of investigating WD planetary systems, it could be especially important. As noted by \cite{FarihiEtAl-2013}, bodies that have sublimated away all their free water can still retain much of the initial water fraction inside hydrated, chemically evolved rocks, subjected to much higher temperatures. Note that hydration reactions are contingent upon having liquid water in the system, which in turn depends on the amount of available heat. Therefore, serpentinization is strongly coupled to both the initial composition and the formation time.

In addition to considering the effects of the main sequence evolution, other aspects of the model of \cite{JuraXu-2010} can be improved:

(d) \emph{Porosity} - objects with radii of up to hundreds of km are believed to have considerable porosity. This is supported by theory \citep{Weidenschilling-1997,WahlbergJohansen-2014}, remote observations \citep{WeissmanLowry-2008,BaerEtaAl-2011}, meteorite samples \citep{ConsolmagnoEtAl-2008,Macke-2010}, and laboratory experiments \citep{Leliwa-KopystynskiEtAl-1994,DurhamEtAl-2005,YasuiArakawa-2009}. \cite{JuraXu-2010} assume that the bulk density of Ceres (2.1 g cm$^{-3}$) represents the typical bulk density of small or intermediate sized icy bodies. However Ceres has a radius of nearly 500 km, it has a semi major axis of only 2.75 AU, and it has been heavily processed thermally and mechanically \citep{NeveuEtAl-2015}. Considering the size range relevant to this paper (up to 100 km in radius, see Section \ref{SS:Configuration}), the bulk density of 1 g cm$^{-3}$ is probably a more appropriate upper limit. To support this claim we have compiled a list (Table \ref{tab:densities}) of all the known solar system objects with a radius between 50-100 km, a semi major axis of approximately 5 AU or more, and that have measured densities (thus limited to binaries or moons).

\begin{table*}
\caption{Densities of small icy objects}
\smallskip
\begin{minipage}{13.5cm}
\centering
\begin{tabular}{|l|l|l|l|l|}
\hline
{\bf Object} & {\bf type} & {\bf Orbit} & {\bf Radius} & {\bf Density} \\ \hline

617 Patroclus & Trojan binary & 5.2 AU & 71/56 km & 0.88 g cm$^{-3}$ \footnote{\cite{MarchisEtAl-2006}} \\ \hline

Amalthea & Jupiter moon & 5.2 AU & 83.5 km & 0.857 g cm$^{-3}$ \footnote{\cite{AndersonEtAl-2005}} \\ \hline

Janus & Saturn moon & 9.5 AU & 89.5 km & 0.63 g cm$^{-3}$ \footnote{\cite{Thomas-2010}} \\

Epimetheus & Saturn moon & 9.5 AU & 58.1 km & 0.64 g cm$^{-3}$ \\ \hline

Teharonhiawako & Kuiper belt binary & 43.8 AU & 89/65 km & 0.6 g cm$^{-3}$ \footnote{\cite{VileniusEtAl-2014}} \\

2001 XR$_{254}$ & Kuiper belt binary & 43.1 AU & 85/70 km & 1 g cm$^{-3}$ \\

2001 QY$_{297}$ & Kuiper belt binary & 43.7 AU & 84/77 km & 0.92 g cm$^{-3}$ \\

Borasisi & Kuiper belt binary & 43.5 AU & 63/51 km & 2.1 g cm$^{-3}$ \\ \hline

\end{tabular}
\label{tab:densities}
\end{minipage}
\end{table*}

With the exception of Borasisi, a very small KBO which has a highly atypical density for its size (more than Pluto, at one tenth of its size), and probably belongs to a special class of KBOs related to past cataclysimic events \citep{MalamudPrialnik-2015,BarrSchwamb-2016}, all other objects are less dense than the upper limit we have set. Among objects with radii less than 50 km, and in particular objects which lose much of their internal water during the RGB and AGB phases, the bulk density should be even lower.

(e) \emph{Heat transport} - treating the thermal conductivity coefficient as a constant is a simplifying compromise that may be improved by using standard comet modeling techniques. \cite{JuraXu-2010} point this out as a caveat, noting that their calculations may overestimate the rate at which heat is conducted into the interior and thus also overestimate the amount of water that is lost. The more accurate way to calculate the effective thermal conductivity is to locally account for both the composition and the porosity, since ice, unaltered rock and serpentinized rock, all have different thermal conductivities \citep{MalamudPrialnik-2015}. Porosity is also of great importance since it considerably decreases the effective thermal conductivity \citep{PrialnikMerk-2008}. In addition to conduction, heat may also be transported by advection. Advection refers to heat transport by any fluid that contains thermal energy, such as water liquid or vapour, of some conserved quantity or material via bulk motion. In other words, the mass fluxes (gas or liquid) themselves carry the internal energy of the moving particles as a byproduct of the motion. Conduction, as well as advection, are the main heat transport mechanisms used for modeling small objects \citep{Prialnik-1992}. 

(f) \emph{Mass transport} - \cite{JuraXu-2010} have developed a thermal evolution model in which they assumed that the ice is instantly transported out of the body as soon as it is vaporized without actually calculating the physical flow. They provide the vaporization temperature, which corresponds to the vapor saturation pressure, given the local hydrostatic pressure by self-gravity (increasing toward the center of the object). However this temperature only reflects the point at which water vapor is in thermodynamic equilibrium with its condensed state. In standard models of comet-like porous structures, ice would actually sublimate at far lower temperatures, and would only condense if the gas pressure inside the pores is locally higher than the saturated vapor pressure. In fact, the vaporization temperature they provide is up to hundreds of degrees higher than even the melting temperature of water. Therefore, realistically, their calculation greatly underestimates the amount of water that should be expelled from the interior.

\subsection{Model description}\label{SS:Description}
In order to alleviate these concerns, we investigate the problem by using a novel model that has been previously applied for studying several solar system icy bodies \citep{MalamudPrialnik-2013,MalamudPrialnik-2015,MalamudPrialnik-2016,MalamudPerets-2016}. This generic model has been developed in order to analyze the evolving thermal, chemical and physical state of icy objects up to several hundred km in radius. In objects with radii of approximately 250 km or more, size becomes a factor, and the model may include related features, such as compaction by self-gravity and maintenance of hydrostatic equilibrium \citep{MalamudPrialnik-2015}. However here we consider much smaller objects (see Section \ref{SS:Configuration}), and thus the code is run in its more basic mode, assuming an initial uniform density profile, in similar configuration to the studies of \cite{PrialnikMerk-2008} and \cite{MalamudPrialnik-2013}. 

In terms of energy sources, the model primarily considers short and long lived radionuclides, latent heat released/absorped by geochemical reactions and surface insolation. The model treats heat transport by conduction and advection. It also follows the transitions among three phases of water (crystalline ice, liquid and vapor). In principle crystallization of initially amorphous ice may also result in some internal activity but it is entirely inconsequential for the survival of water during the RGB and AGB phases, since it contributes relatively little energy. We therefore ignore amorphous ice by choosing an initial composition of fully crystalline ice. The model consistently considers the contribution/absorption of energy by phase transitions among the various phases of water (condensation/deposition/freezing/sublimation/evaporation/melting), noting that when these contributions enter the total energy budget they are actually of minor significance compared to other internal heat sources. The model also follows the transitions among two phases of silicates (aqueously altered rock and non-altered rock), accounting for thermal (conductivity, heat capacity) and physical (density/porosity) changes in the solid phases, on top of the energy contribution (latent heat release/absorption). Note that if the internal temperature becomes sufficiently high (typically above 670 K), the rock may undergo the reverse process of serpentinization, known as dehydration, in which the rock exudes the water it had previously absorbed, and releases it back into the system. For our choice of model parameters this only occurs in one extreme (end) case, and even in this case dehydration is shown to have a marginal effect, as we shall demonstrate in Section \ref{SS:Evolution}. 

\subsection{Governing equations and numerical scheme}\label{SS:Equations}
We have six different components that we denote by subscripts: $u$ - aqueously unaltered rock; $p$ - aqueously processed rock; $c$ - crystalline water ice; $\ell$ - liquid water; $v$ - water vapor.

The independent variables are: the cumulative volume $V$; temperature $T$; densities $\rho_w=\rho_c+\rho_{\ell}$, $\rho_v$ and $\rho_d$, as well as the mass fluxes $J_v$ (water vapor) and $J_\ell$ (liquid water), as functions of 1-D space and time $t$. The set of equations to be solved is:

\begin{align}
\label{eq1}
\frac{\di(\rho U)}{\di t} + \frac{\di}{\di V}\left(-K\frac{\di T}{\di V}\right)+\frac{\di(U_vJ_v + U_\ell J_\ell)}{\di V}+q_\ell\Hcal_\ell-S=0 \\
\label{eq2}
\frac{\di \rho_v}{\di t} + \frac{\di(J_v)}{\di V}  = q_v \\
\label{eq3}
\frac{\di \rho_w}{\di t} + \frac{\di J_\ell}{\di V}  = - q_v + \frac{2A_w}{A_u} \left(R_D \rho_p - R_S \rho_u\right) \\
\label{eq4}
J_v = - \phi_v \frac{\di\left(P_v / \sqrt{T}\right)}{\di V}  \\
\label{eq5}
J_\ell = -\phi_\ell \left(\frac{\di(P_\ell)}{\di V}+\rho_\ell g\right) \\
\label{eq6}
V(x,t)=V_s(t)(1-q^{x-c}), \qquad q<1
\end{align}

Equations (\ref{eq2}-\ref{eq3}) are the mass conservation equations, where $R_S$ is the serpentinization rate, $R_D$ is the dehydration rate, $A_u$ and $A_w$ denote the molecular weights of water and unaltered rock respectively and $q_v$ is the rate of sublimation/evaporation or deposition/condensation, respectively. The mass fluxes are given by equations (\ref{eq4}-\ref{eq5}), where $\phi_v$ and $\phi_\ell$ are the permeability coefficients, $P_v$ and $P_\ell$ are the gas and liquid pressures respectively and g the local gravitational acceleration (although in small objects this term is negligible). 

In the energy conservation equation (\ref{eq1}), $U$ denotes energy per unit mass, $\Hcal_{\ell}$ is the latent heat of fusion, $q_{\ell}$ is the rate of melting/freezing, and $K$ is the effective thermal conductivity, accounting for heat transferred by conduction, while $(U_vJ_v + U_\ell J_\ell)$ accounts for the heat transferred by advection. The sum of all energy sources $S$ includes  the energy lost by sublimation, and all the other possible internal heat sources, such as radiogenic heating and heat released or absorbed by geochemical reactions. The last equation, eq. (\ref{eq6}), defines the volume distribution (grid zoning) over the range of $x[c,s]$ (center - surface) through a geometric series. In the present calculations we take $q=0.95$, since fine zoning close to the surface is not required. All the other variables are easily derived from the independent variables and the volume delineation. The boundary conditions adopted here are straightforward: vanishing fluxes at the center and vanishing pressures at the surface. The surface heat flux is given by the balance between solar irradiation (albedo dependent), thermal emission and heat absorbed in surface sublimation of ice. We note that for an eccentric orbit, distance variations on the scale of the orbital period, would render the changes in the surface boundary condition extremely fast, and therefore the calculation extremely slow, as time steps are dynamically adjusted. For relatively small eccentricities one may circumvent this problem by considering an effective circular semi-major axis, producing an equivalent average insolation \citep{WilliamsPollard-2013}. Since this technique typically introduces very small corrections to the semi-major axis, we simply assume a circular motion, so that the semi-major axis is our single orbital free parameter.

The model uses an adaptive-grid technique, specifically tailored for objects that change in mass or volume. In this study we model relatively small objects, and therefore their size remains constant throughout the evolution, or (in the largest objects) it remains nearly constant, however the porosity profile always changes due to the internal migration and redistribution of water. The numerical solution is obtained by replacing the non-linear partial differential equations with a fully implicit difference scheme and solving a two-boundary value problem by relaxation in an iterative process. Time steps are adjusted dynamically according to the number of iterations. The volume is distributed over a variable $x$ that assumes integer values of $i$, from $i=1$ at the center to $i=I$ at the surface; here we use $I=100$. Since temporal derivatives are taken at constant $V$, whereas $V=V(x,t)$, the following transformation is implemented in the difference scheme: 

\begin{equation}
\left(\frac{\di}{\di t}\right)_V = \left(\frac{\di}{\di t}\right)_x - \left(\frac{\di V}{\di t}\right)_x \ . 
\left(\frac{\di}{\di V}\right)_t
\end{equation} 

Apart for the governing equations, the full details concerning all other constituent relations can be found in earlier papers \citep{PrialnikMerk-2008,MalamudPrialnik-2013,MalamudPrialnik-2015,MalamudPrialnik-2016}.

\section{Results}\label{S:Results}
\subsection{Model configuration and parameters}\label{SS:Configuration}
As mentioned in Section \ref{SS:Description}, the outcome of the RGB and AGB stellar evolution phases in terms of water retention in the WD's planetary system, depends on several different parameters. We investigate four critical variables: radius, orbital distance, formation time and initial rock/ice mass ratio. The two latter parameters have not been taken into account in any study prior to this. 

(1) Object radius - we consider the following radii: 1, 5, 25, 50 and 100 km. This covers the entire size spectrum from small comets to moonlet sized objects. Objects with larger radii are not considered for a couple of reasons. First, assuming an initial power law size distribution \citep{Johansenl-2015}, larger objects are less numerous in quantity by up to several orders of magnitude, and are therefore much less likely to trigger accretion events. Second, we determine that for a radius of 100 km, the object is sufficiently large in order to always retain some fraction of the initial water content during the post main sequence evolution, that is, for every combination of the other parameters (see Section \ref{SS:Evolution}). Larger, moon-sized objects are therefore guaranteed to do the same. Truly sizable objects such as super-moons or super-Earths are so large that even their atmospheres will not be completely eroded \citep{RamirezKaltenegger-2016}, let alone their interior. 

(2) Orbital distance - according to \cite{KunitomoEtAl-2011} and \cite{VillaverEtAl-2014} any object at a minimal distance of less than a few AU, is in danger of being completely engulfed or otherwise destroyed by the expanding envelope of the post main sequence star. This minimal initial distance is extended for small objects, but only for them, as a result of a different consideration. According to \cite{VerasEtAl-2014a}, all objects with radii of less than 10 km are not likely to survive the post main sequence evolution if their initial orbital distance is less than about 7 AU due to spin acceleration by the YORP effect leading to their breakup after crossing the critical spin limit. For objects with radii between 1-5 km this breakup distance could extend even further to a few tens of AU. For the larger objects, we show in Section \ref{SS:Evolution} that at an initial orbital distance of 5 AU, an object with a (maximal, in our sample) radius of 100 km may lose all of its \emph{free} water during the RGB and AGB phases (if serpentinization occurs, a fraction of the water would be retained in the rock). Our minimal initial orbital distance is therefore 5 AU, and it is consistent with the minimal distance chosen by \cite{JuraXu-2010}. To set up the rest of the grid we consider the orbital distances of 7, 10, 20, 30, 40 and 75 AU. Beyond 75 AU our results indicate that full water retention is insured even for a 1 km radius object. Note that the initial orbital distance increases, primarily during the AGB phase, considering conservation of angular momentum.

(3) Formation time - the formation time as a function of distance, that is, the time it takes the object to reach its full initial size in orbit around the young star, is an unconstrained parameter, even in our own solar system. As a rule, the shorter it is, the higher the temperatures attained by the object during the first few Myr of its evolution, noting that peak temperatures also greatly depend on size. For instance, a 1 km radius object cools so efficiently that it is effectively insensitive to the choice of this parameter, however the sensitivity increases with size. In Section \ref{SS:Description} we discuss the lower and upper limits for the formation time, for the minimum orbital distance of 5 AU, which is 2-5 Myr. At larger distances the formation time obviously increases, however our results indicate that for the maximal radius of 100 km, the evolution is actually insensitive to any formation time greater than about 5 My. Hence, we set the formation time of 5 Myr as our upper limit. We further investigate shorter formation times: 3 Myr and 4 Myr. At 2 My, the peak temperature during the early evolution could be so high, that if the rock has underwent serpentinization, it could also experience vast dehydration, and as a result water would not survive in the rock. We set up the grid such that even in the end case (maximum size, and minimum orbital distance and formation time) the amount of dehydration is marginal, since we are mainly interested in the maximal amount of water that can survive in the rock. Hence, our formation times are 3, 4 and 5 Myr.

(4) Initial rock/ice mass ratio - this ratio initially depends on the location of the object as it forms in the protostar nebula, and like the formation time this parameter is unconstrained. In comets, the typical ratio observed in the comma usually varies between 1/3 and 3, however this measurement is biased since the mass of the dust ejected from comets is mainly a function of the flow of gas escaping from the surface and also the size of the grains. As a heuristic approach studies usually assume a ratio of 1. This estimation is also in reasonable agreement with theoretical studies \citep{MarboeufEtAl-2014}. The more recent encounter with comet P67/Churyumov–Gerasimenko hints that many comets might actually be richer in silicate content than previously believed, with a ratio of approximately 4 \citep{PatzoldEtAl-2016}. A recent study regarding the possible structure and composition of KBOs also suggests a typical Kuiper belt ratio of about 3 \citep{MalamudPrialnik-2015}. We therefore finalize our grid by considering three initial rock/ice mass ratios: 1, 2 and 3 (that is, a rock mass fraction of 50\%, 67\% and 75\% respectively).

The total number of possible models is thus 315, determined by the number of variable parameters (5 x 7 x 3 x 3). Our data set accordingly consists of 315 individual production runs. All other model parameters are equal, and the main ones are listed in table \ref{tab:param}. Note that the albedo is not constant because water-rich surfaces are more reflective. The albedo-composition relation is not constrained in the literature, and therefore we assume a simple linear law. For surfaces similar to comets with an insulating dust layer the albedo is 0.04, whereas completely icy surfaces have a high albedo comparable to Enceladus.  

\begin{table*}
\caption{Initial and physical parameters}
\centering
\smallskip
\begin{minipage}{14.2cm}
\begin{tabular}{|l|l|l|}
\hline
{\bf Parameter}                      & {\bf Symbol} & {\bf Value} \\ \hline
Initial uniform temperature          & $T_0 (t=0)$ & 70 K \\
Nominal $^{235}$U abundance          & $X_0$($^{235}$U) & $6.16\cdot 10^{-9}$ \\
Nominal $^{40}$K abundance           & $X_0$($^{40}$K) & $1.13\cdot 10^{-6}$ \\
Nominal $^{238}$U abundance          & $X_0$($^{238}$U) & $2.18\cdot 10^{-8}$ \\
Nominal $^{232}$Th abundance         & $X_0$($^{232}$Th) & $5.52\cdot 10^{-8}$\\
\hline
Albedo                               & $\Acal$ & 0.04+0.8$f_s$ (surface water fraction)\\
Ice specific density                 & $\varrho_c$ & 0.917 g cm$^{-3}$ \\
Water specific density               & $\varrho_\ell$  & 0.997 g cm$^{-3}$ \\
Rock specific density (u)            & $\varrho_u$ &  $3.5$ g cm$^{-3}$ \\
Rock specific density (p)            & $\varrho_p$ &  $2.9$ g cm$^{-3}$ \\
Water thermal conductivity           & $K_\ell$ & $5.5\cdot 10^4$ erg cm$^{-1}$ s$^{-1}$ K$^{-1}$\\
Ice thermal conductivity (c)         & $K_c$ & $5.67\cdot 10^7/T$ erg cm$^{-1}$ s$^{-1}$ K$^{-1}$\\
                                     &       & erg cm$^{-1}$ s$^{-1}$ K$^{-1}$\\
Rock thermal conductivity (u)        & $K_u$& $10^5/(0.11+3.18\cdot 10^{-4}T)+$\\
                                     &       &$3.1\cdot 10^{-5}T^3$ erg cm$^{-1}$ s$^{-1}$ K$^{-1}$\\
Rock thermal conductivity (p)        & $K_p$& $10^5/(0.427+1.1\cdot 10^{-4}T)+$\\
                                     &       &$8.5\cdot 10^{-6}T^3$ erg cm$^{-1}$ s$^{-1}$ K$^{-1}$\\
\hline
\end{tabular}
\label{tab:param}
\newline Sources: $X_0$ \citep{Prialnik-2000}; $\varrho_u$ \citep{WattAhrens-1986}; $\varrho_p$ \citep{TyburczyEtAl-1991}; $K_{\ell,c,a}$ - \cite{MalamudPrialnik-2013}; $K_{u,p}$ - \cite{MalamudPrialnik-2015}.
\end{minipage}
\end{table*}

\begin{figure}
\begin{center}
\subfigure[Luminosity as a function of time.] {\label{fig:LvsTime}\includegraphics[scale=0.52]{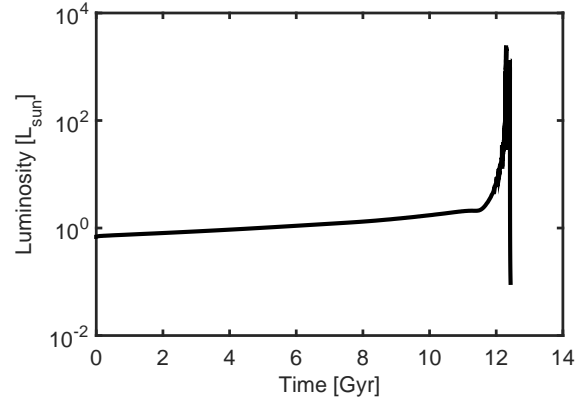}}
\subfigure[Mass as a function of time.] {\label{fig:MvsTime}\includegraphics[scale=0.52]{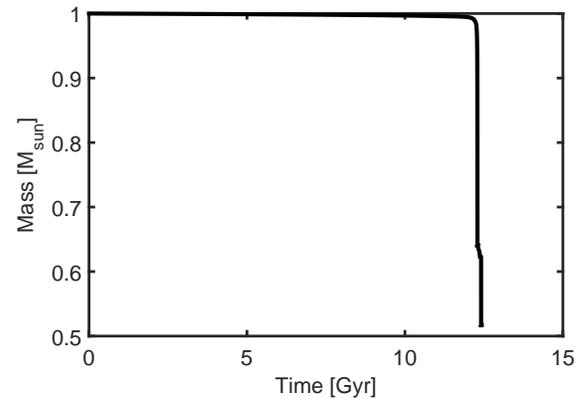}}
\end{center}
\caption{MESA evolution of a 1 M$_{\odot}$ star with solar metalicity. On the left we plot the significant increase in star luminosity after the main sequence. On the right we plot stellar mass loss, primarily during the AGB phase.}
\label{fig:MESAevolution}
\end{figure}

The star's luminosity changes as a function of time, as well as its mass, and both are obtained via the MESA stellar evolution code \citep{PaxtonEtAl-2011}, for a 1 M$_\odot$ star with solar metalicity. Unlike \cite{JuraXu-2010} we do not consider the evolution of more massive stars, since the number of free parameters in this study is already substantial (note that different combinations of star masses and metalicities yield numerous post main sequence evolutions, varying in luminosity and duration). Investigation of more massive stars is left for future studies. For our choice of star mass and metalicity, the luminosity as a function of time is shown in Panel \ref{fig:LvsTime}. The total duration of the evolution, including the main sequence, RGB and AGB phases is 12.43 Gyr. The evolution is truncated after the AGB phase, when the initially 1 M$_\odot$ star shrinks to about 0.517 M$_\odot$, as shown in Panel \ref{fig:MvsTime}. As the star losses mass, conservation of angular momentum dictates an orbital expansion. The typical run time of each model is several hours on a single 2.60GHz, Intel CPU.

Since it is impossible to show the detailed evolutions of 315 distinct models, we will focus in Section \ref{SS:Evolution} on one particular evolutionary path. The model we have chosen to show represents the end case in our sample, that is, the largest radius (100 km), the smallest orbital distance (5 AU), shortest formation time (3 Myr), and highest rock/ice mass ratio (3). This combination of parameters results in the warmest early evolution possible, and therefore this evolutionary path demonstrates all of the important aspects of our model: (a) early melting, differentiation and stratification; (b) rock phase transitions from anhydrous to hydrous rock and then the inverse reaction; (c) water ablation during the post main sequence, proceeding from the surface inward. 

We then continue to show the results for the entire planetary system in Section \ref{SS:Retention}, incorporating all 315 models. In this case however we will not present full evolutions, and instead concentrate on the final fraction of water that survives the RGB and AGB phases.

\subsection{The evolutionary course}\label{SS:Evolution}
\begin{figure}
\begin{center}
\subfigure[Warm early evolution.] {\label{fig:ROWbeginning}\includegraphics[scale=0.52]{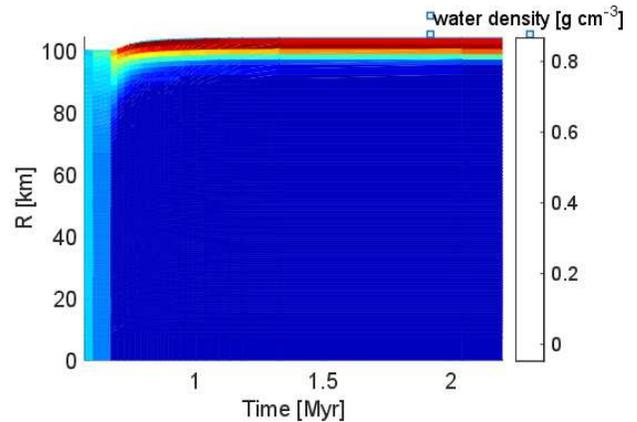}}
\subfigure[Post main sequence water ablation.] {\label{fig:ROWend}\includegraphics[scale=0.52]{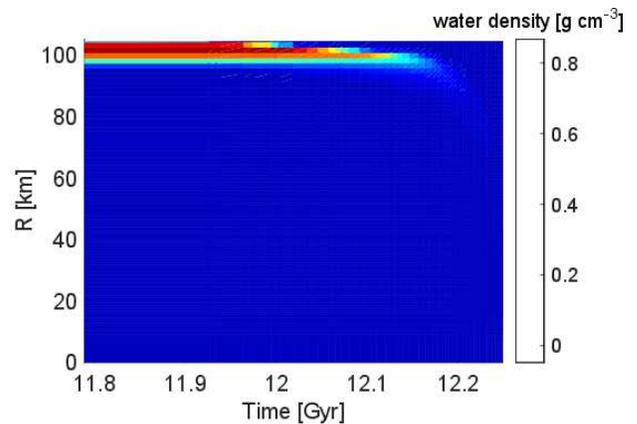}}
\end{center}
\caption{A surface plot of water density as a function of time and radial distance from the center of the minor planet. On the left we show how the originally homogeneous structure becomes stratified following the water/rock differentiation process. On the right we show how water is undergoing ablation near the surface during the last few Gyr of the evolution.}
\label{fig:WaterDensity}
\end{figure}

As mentioned in Section \ref{SS:Configuration}, here we present one evolution calculation for a body that has a radius of 100 km, it is 75\% rock by mass, it formed only 3 My after the formation of CAIs, its initial orbital distance is 5 AU and its final orbital distance after stellar mass loss is 9.67 AU. The course of the evolution may be illustrated by various surface plots showing properties as a function of time and radial distance from the center of the body, such as the temperature, total density, specific densities, porosity, mass fluxes, energy sources, etc. Here we shall focus on the water density, shown in Figure \ref{fig:WaterDensity}, since it contains all the relevant information needed in order to understand the main aspects of the model.

In Panel \ref{fig:ROWbeginning} we show the early evolution. The x-axis shows the time interval, ranging from 0.6 Myr to 2.2 Myr. The y-axis shows the radial distance from the center of the body. Note that the upper boundary of the y-axis changes with time, from 100 km to a maximum of about 104 km. This marginal change is caused by the differentiation process of the body. As water migrates to freeze near the surface, forming the outer icy crust, this entails a small expansion in volume when the porosity eventually reduces to zero and the outer layers must expand under pressure in order to compensate. In all other evolutionary models the radial expansion is either smaller or zero. The z-axis, depicted by the color scale in the right side of the panel, shows the internal distribution of water. The density changes from a minimal value of 0, to a maximum value that approaches the specific density (zero-porosity density) of water ice, 0.917 g cm$^{-3}$. Before 0.6 Myr, water is still uniformly distributed, and the body is completely homogeneous. After the temperature rises to above the melting point of water the body rapidly differentiates. By 1.3 Myr, the inner core is entirely depleted of water. Water has instead migrated to the cold outer layers or else absorbed by serpentinization in the rock. The ice that froze in the outer crust has lowered its porosity considerably. This configuration is maintained for approximately 12 Gyr, as shown in Panel \ref{fig:ROWend}, however when the surface temperature begins to rise, sublimation starts, triggering the ablation of water from the crust. The heat provided by external insolation at merely 5 AU, penetrates deeper with time, until by 12.2 Gyr virtually all of the free water in the system is expelled from the body. This occurs approximately 200 Myr prior to the end of the AGB evolution. This immediately emphasizes the importance of the early evolution (formation time), as well as the choice of initial composition (rock/ice mass ratio), because had the body remained completely homogeneous, or if the crust was thicker, it would have taken longer for the heat to penetrate and remove all the free water.

\begin{figure}
\begin{center}
\subfigure[Evolution of rock.] {\label{fig:RODUcenter}\includegraphics[scale=0.52]{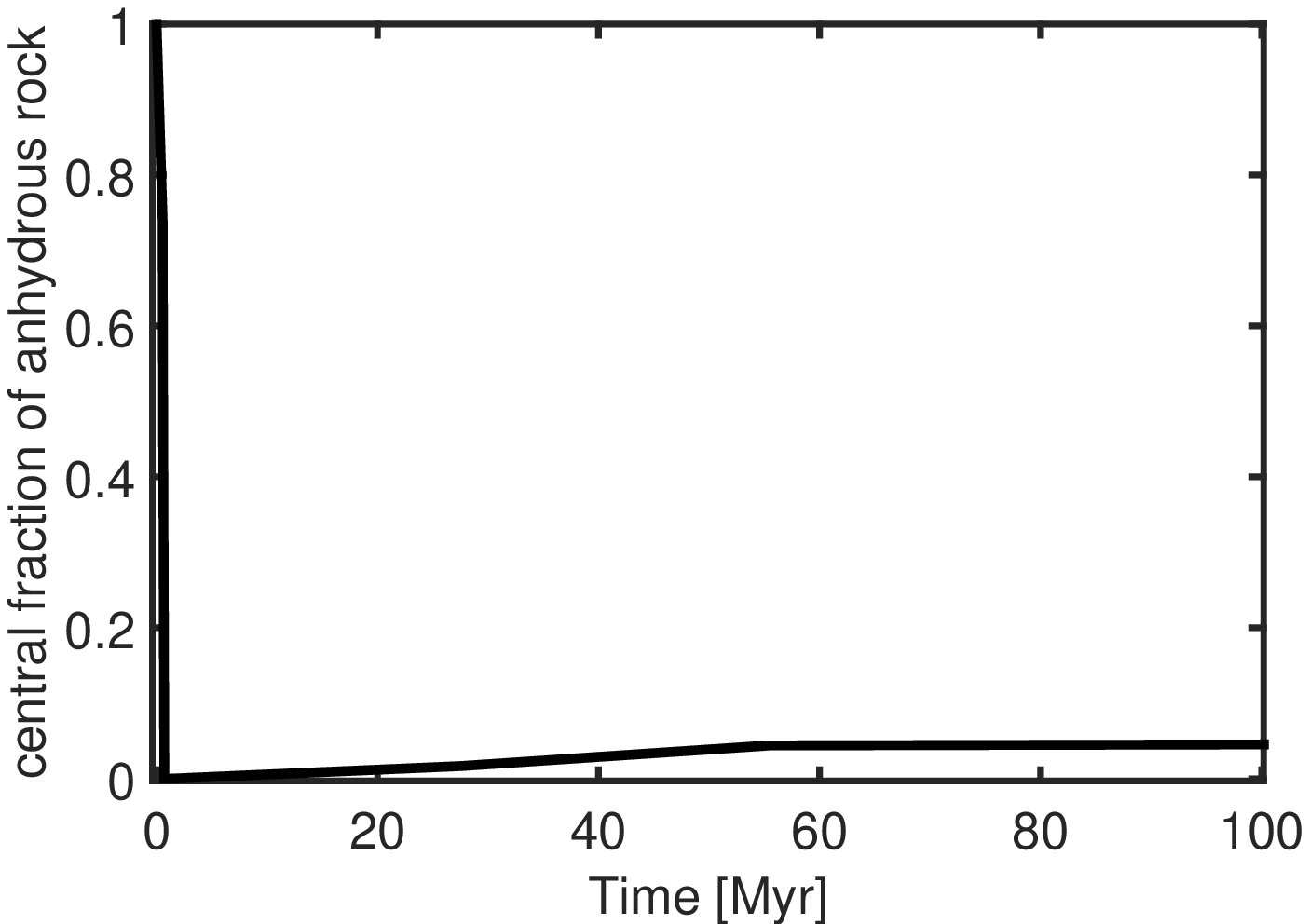}}
\subfigure[Evolution of temperature.] {\label{fig:Tcenter}\includegraphics[scale=0.52]{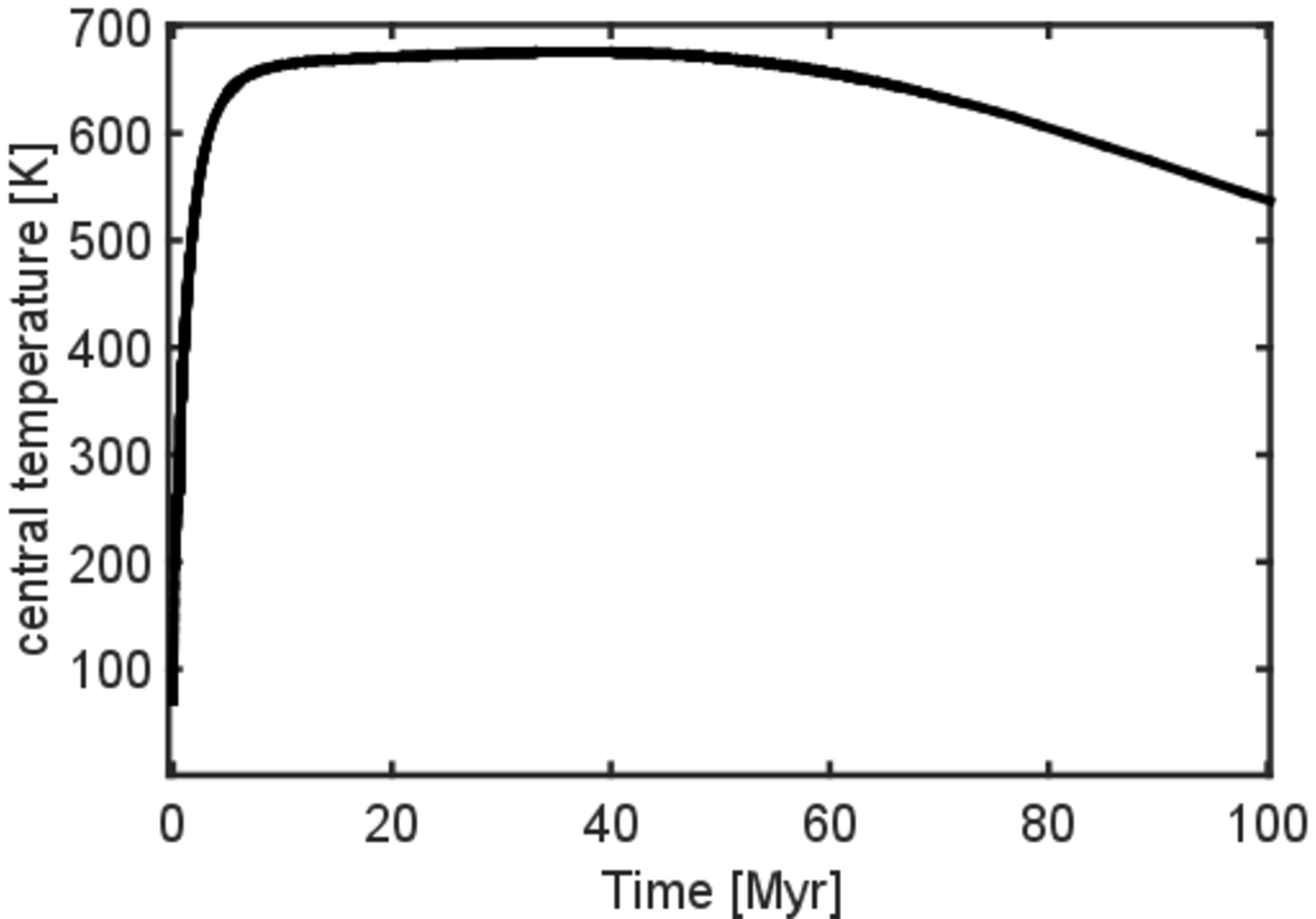}}
\end{center}
\caption{The rock phase transitions on the left indicate first serpentinization that occurs after the temperature reaches 273K and liquid water is available, and then rock dehydration when peak temperatures reach 670K. The corresponding temperatures as a function of time appear on the right.}
\label{fig:PhaseTransfer}
\end{figure}

Although in this case we have shown that all the free water in the system has sublimated away, a significant mass fraction of the now hydrated rock, about 13\%, is actually attributed to the absorption of water. This can be seen in Figure \ref{fig:PhaseTransfer}, which shows the progression of rock phase transitions during the first 100 Myr, in the central part of the body. Panel \ref{fig:RODUcenter} shows the fraction of anhydrous rock, 1 denoting completely anhydrous rock. As the temperature increases (in Panel \ref{fig:Tcenter}), ice turns to liquid water, which then starts reacting with the rock, while migrating outward as shown in Figure \ref{fig:WaterDensity}. Accordingly, the fraction of anhydrous rock drops steeply from 1 to 0. This implies that after a short early evolution almost all the rock has been serpentinized, that is, it contains water on the molecular level. The temperature in the center continues to increase, finally reaching a peak of about 675 K. During this time, the rate of dehydration reactions increases and about 5\% of the rock dehydrates before the temperature decreases again. In order to undergo full dehydration, the object must be larger than 100 km, or form quicker. Since we have limited our parameter grid, this analysis is beyond the scope of the paper.

\subsection{Water retention}\label{SS:Retention}
\begin{figure*}
\begin{center}
\subfigure[Total water fraction.] {\label{fig:total}\includegraphics[scale=0.5]{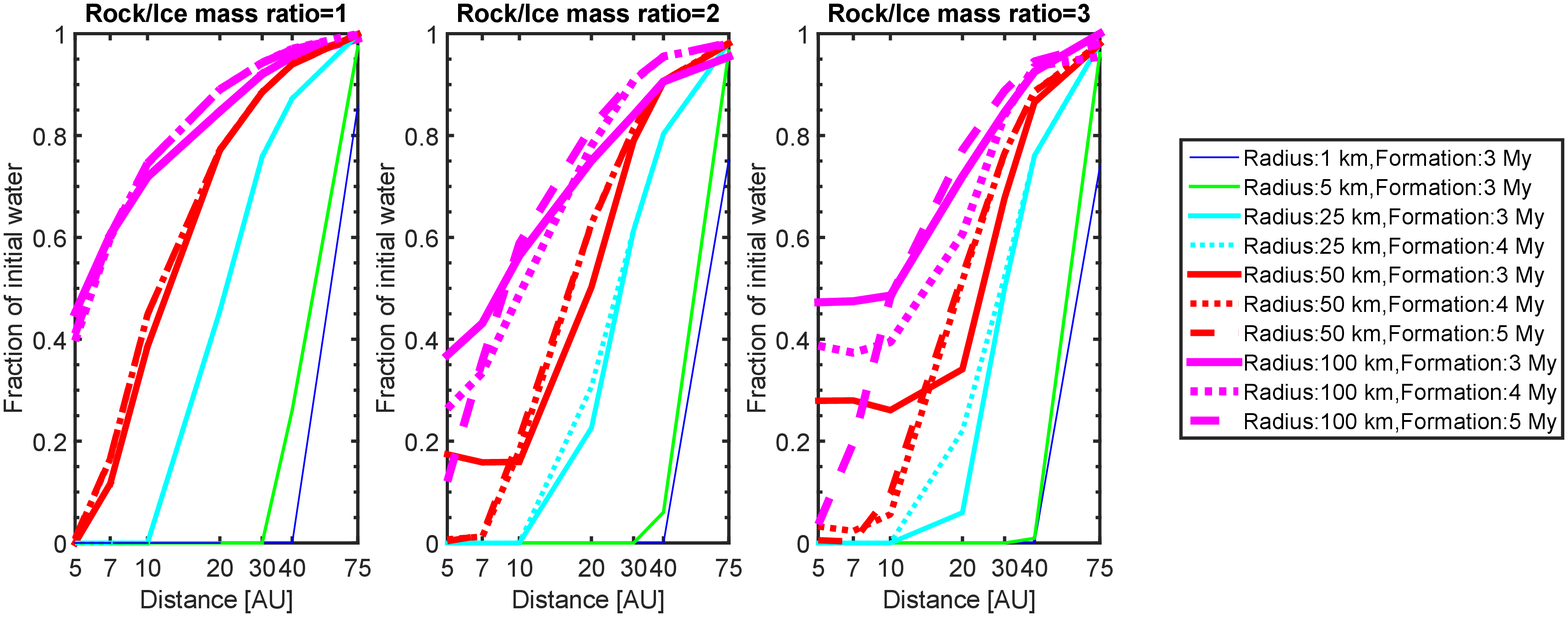}}
\subfigure[Free water fraction.] {\label{fig:free}\includegraphics[scale=0.5]{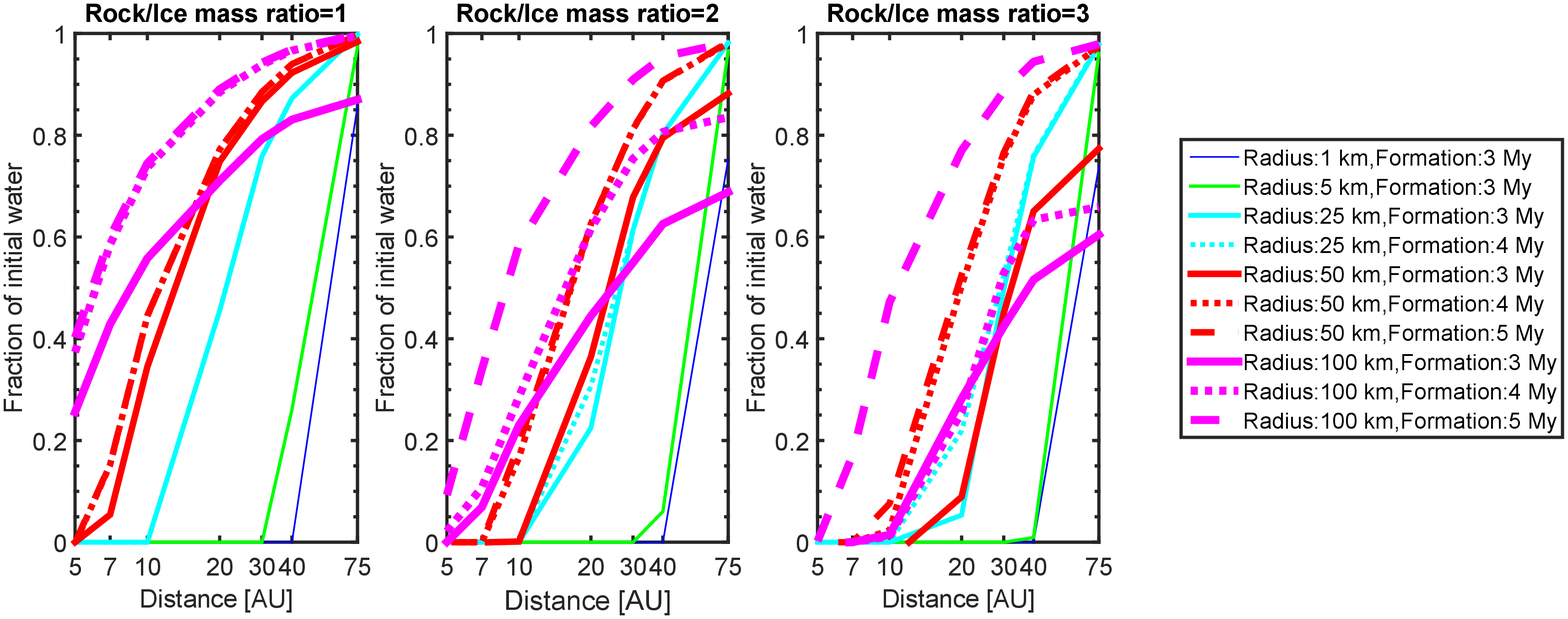}}
\end{center}
\caption{Bulk water fraction in the system. The retention of water is shown as a function of all model parameters. \emph{Total water fraction} includes water ice + water embedded in the rock via hydration. \emph{Free water fraction} includes water ice only.}
\label{fig:WaterFraction}
\end{figure*}

In this section we discuss the bulk amount of water surviving in the planetary system, as a function of our free model parameters. Figure \ref{fig:WaterFraction} shows the final fraction of water, based on the end state of all the models after the AGB stellar evolution phase. The difference between Panels \ref{fig:total} and \ref{fig:free} is that the former shows the total fraction of the initial water, including water that were embedded in hydrated rocks, whereas the latter refers only to the fraction of free water. Each panel consists of three subplots, each representing a different choice for the initial composition. Within each subplot there are multiple lines, depicting the final water fraction as a function of the initial orbital distance. Each line is characterized by a specific color and width, as well as a style. The line width increases with the size of the object, so thick lines represent large objects, and each line style corresponds to a different formation time. 

Panel \ref{fig:total} summarizes the main results of the paper. Comet-like objects that are 1-5 km in radius are expected to retain some if not all of their water, only if their initial orbital distance is beyond the orbit of Neptune. This result is actually in good agreement with the model of \cite{VerasEtAl-2014a}, which also predicts YORP critical spin breakup if the initial orbit of a comet is less than a few tens of AU, so both models validate each other, obtaining the same result for different reasons. For larger objects, the general trend is that water can survive increasingly closer to the star, with increasing size, however there are small differences.

For example, note the rightmost subplot in Panel \ref{fig:total} with a rock/ice mass ratio of 3. For a 50 km radius object, the fraction of water that can survive between 5 AU and about 15 AU is very small or zero, depending on the distance. However the curve that corresponds to a formation time of only 3 My is very different. It shows that the total water fraction remains an almost constant 0.3 for this entire distance range. This happens because the interior temperatures attained in this case, and only in this case, can be higher than the melting temperature of water. Therefore some of the anhydrous rock reacts with liquid water, embedding it in the hydrated rock. In fact, the corresponding rightmost subplot in Panel \ref{fig:free} clearly shows that for a formation time of 3 Myr, the fraction of free water is actually lower compared to the other formation times, given that in this case the body differentiates, and the resulting structure is stratified. Hence, the water concentrates in the outer crust and is easier (quicker) to remove. Yet despite the fact that there is actually less free water in the body, the total fraction of surviving water is larger. For a rock/ice mass ratio of only 2 (center subplots), the analysis for a 50 km radius object is almost identical. The difference is that the smaller initial rock fraction, results in lower peak temperatures, hence less anhydrous rock is serpentized, the total water fraction equals only about 0.2 and the turnoff point is at 10 AU instead of 15 AU. 

Another interesting result is that for a rock/ice mass ratio of 1 (or less), the final water fraction is completely insensitive to the choice of the formation time parameter. The reason is that the peak temperatures during the main sequence are relatively low. However this would certainly change if the parameter grid was extended to include shorter formation times or larger object radii.

\section{Discussion}\label{S:discussion}
The results in Section \ref{S:Results} indicate that an extrasolar Kuiper belt around a G-type star must retain most of its water following the luminous RGB and AGB phases, and that only comet sized icy objects are sensitive to significant water loss, depending on their exact distance. At closer distances, the amount of water surviving is strongly coupled to the size of the object, but also to other parameters. We have shown here for the first time that the formation time of the object as well as its assumed initial composition are also important. At an initial orbital distance of only 5 AU, an object with a 100 km radius is sufficiently large to retain up to 50\% of its internal water. By extrapolation, larger objects will of course retain an even bigger fraction. At the same orbital distance, we have also shown that under no circumstances, can an object with a 50 km radius (or smaller) retain free water in its interior, however in some cases water may still survive, embedded in the interior rocks through chemical reactions. Compared to the results reported here, the model of \cite{SternEtAl-1990} greatly overestimates the amount of water loss in comet-like objects, whereas the model of \cite{JuraXu-2010} considerably underestimates water loss in general, and in particular for big objects. 

Estimating precisely the bulk amount of water that survives in planetary systems as a function of the initial orbital distance requires a statistical analysis. This is a complicated task, that is beyond the scope of this paper. Several assumptions are required, including: the size distribution of icy objects, the formation time as a function of orbital distance and the initial composition as a function of orbital distance. We may nevertheless conclude, even without a detailed statistical analysis, that if water is initially present, it can certainly survive in the WD system, in a variety of circumstances and locations. It is thus difficult to reconcile water retention on the one hand, and the scarcity of observations of ongoing accretion events of water-bearing objects on the other hand. In the following subsection \ref{SS:scarcity} we contemplate the likelihood of several potential explanations for this problem.

\subsection{Scarcity of water detections}\label{SS:scarcity}

\subsubsection{Insufficient high resolution observations}
The number of WDs cataloged as metal polluted \citep{DufourEtAl-2007,KleinmanEtAl-2013,KeplerEtAl-2015,KeplerEtAl-2016} is rapidly increasing and may now amount to thousands. However the highest spectral resolution is obtained for relatively nearby WDs using Keck, VLT and HST \citep{KleinEtAl-2010,KleinEtAl-2011,GansickeEtAl-2012,KoesterEtAl-2014}, yielding high resolution observations for several tens of WDs. Although more future high resolution observations are certainly required, this sample appears sufficiently large to conclude that observations of ongoing water-bearing polluting minor planets are indeed rare, and not the result of small statistics. 

\subsubsection{Exo-solar water less abundant}
Several lines of theoretical and observational arguments are in strong support that the majority of exo-solar systems are similar to our own in terms of water content. Infrared observations for example, are teaching us that interstellar ices are commonplace and carry a large fraction of the solid mass in protostellar environments \citep{ObergEtAl-2011}. It is not yet clear what is the degree to which volatiles are inherited from protostellar molecular clouds, or the degree to which their chemistry is being reset during the formation of planetary systems. Even in the latter case, theory predicts that various carbon, nitrogen and oxygen volatiles account for much of the condensible mass that eventually forms planetary bodies. Oxygen is generally the least depleted element among these, and it is typically more common in volatile molecules relative to refractory molecules, by a factor of a few, the most abundant molecule being H$_2$O \citep{PontoppidanEtAl-2014}. 

\subsubsection{Dry bodies are preferentially perturbed to tidal disruption orbits}
Our results indicate that inner system objects are intrinsically dry, or, if sufficiently large, they are relatively drier than in the outer portions of exo-planetary systems. Thus, if minor planets around WDs are depopulated from the inside out, as some models predict, this could principally lead to more detections of dry polluting material, since there is an observational bias for the detection of oxygen in warmer, younger WDs \citep{JuraYoung-2014}. However, it remains to be determined if this sequence of depopulation is categorically correct. It is currently not fully understood exactly how minor planets are delivered to tidal crossing orbits. There are several possibilities: a planetary mean motion resonance between a single giant planet and a belt of planetesimals \citep{DebesEtAl-2012}, however the belt must be massive to be sustained for a long time; direct scattering by a single eccentric planet \citep{BonsorEtAl-2011,FrewenHansen-2014}, however stellar flybys are required in order to scatter the planet's orbit itself and reset the dynamics in the system for a continual flux; multiplanet systems with two planets \citep{DebesSigurdsson-2002}, three planets \citep{MustillEtAl-2014} or more \citep{VerasGansicke-2015,VerasEtAl-2016}, that in some cases enable richer dynamics and can sustain delivery over longer time spans; the ejection of moons into tidal disruption orbits through planet-planet scattering events \citep{PayneEtAl-2016}, however the total number of exo-moons is likely several orders of magnitudes smaller compared to other planetesimal populations; the engulfment of a compact planetary system (2-5 AU) by a star evolving to become a WD leads to orbit destabilization of planetesimals (10-40 AU) through the Lidov-Kozai mechanism triggered by a stellar binary companion, even after several Gyr \citep{PetrovichMunoz-2016}. 

There are three additional possibilities that primarily apply to young WDs: systems with a binary stellar companion, in which a planet may be perturbed following mass-loss induced instability \citep{KratterPerets-2012} or through mass-loss induced Lidov-Kozai oscillations \citep{PeretsKratter-2012,ShapeeThompson-2013,MichaelyPerets-2014,HamersPortegiesZwart-2016}, or, comets perturbed to low angular momentum orbits by anisotropic mass-loss induced natal kicks \citep{StoneEtAl-2015}. In conclusion, the sequence of depopulation is highly uncertain. Clearly more studies are required in order to discern between these, as well as other possibilities. Also, these models depend on the assumed planetary configuration, however despite the great advances in exo-planet research and characterization, our knowledge regarding planetary configurations is currently limited. 

Independently of the depopulation sequence, we also note that in close orbital proximity, only objects that have a radius of 50-100 km (or more) will be able to retain water during the RGB and AGB phases (see Figure \ref{fig:total}). This means that even objects that are supposed to be in the 'dry' zone, may still contain a considerable water fraction. The lack of significant excess oxygen in most detections may then be attributed simply to the size distribution of inner-system planetesimals (the large ones being more rare, however being able to retain water). This explanation is plausible and complementary to the depopulation sequence explanation, however it requires a further statistical analysis, as well as more information about the size distribution in exo-solar planetary systems.

\subsubsection{Water rapidly sublimates and accretes prior to the formation and accretion of the remaining silicate disk}\label{sss:DiffAccretion}
Here we suggest a potential path to avoid water detections in H-dominated WDs, but only for them, if the two main compositional constituents of minor planets, water and silicates, experience a different accretion sequence. Assuming that the perturbed minor planet has a periastron inside or close to the tidal disruption radius, we will show that it should experience rapid water sublimation and subsequent accretion onto the WD. Meanwhile, the rest of disk, composed of the remaining silicate material, would still be in the process of forming and circularizing. Thus, depending on the exact rate of accretion of the ensuing water gas, water detections in H-dominated WD atmospheres, that have a very short sinking time, could be very rare. This explanation does not apply to He-dominated WDs (unless they are extremely young), because such atmospheres can preserve the record of previous water accretion events for comparable or even longer time periods compared to the dissociation and circularization time of the silicate disk, as we discuss below.

According to the analysis of \cite{VerasEtAl-2014b}, a tidally disrupted body disassociates into a highly eccentric ring of fragments from the original body, a process which lasts several tens to hundreds of orbits, and up to 10$^5$ yr depending on the semi-major axis and other parameters. The ring then undergoes subsequent circularization to within the roche radius (several tens of WD radii) via Poynting-Robertson (PR) drag in approximately $10^2$-$10^6$ yr, depending on the initial semi-major axis, the WD cooling age and the size of the disassociated particles \citep{VerasEtAl-2015}. However if the disrupted body has water in it, we suggest that water in the disassociated fragments (i.e., cometesimals) could sublimate rapidly, in similarity to solar system comets. The ensuing gas will then undergo accretion. These processes could be extremely rapid in comparison to the formation and accretion of the remaining silicate disk.

Solar system comets, are known to exhaust their volatiles extremely fast. Jupiter family comets are active over a lifetime of about $10^4$ years or $\sim$1,000 orbits whereas long-period comets fade much faster. Only 10\% of the long-period comets survive more than 50 perihelion passages and only 1\% of them survive more than 2,000 passages \citep{WhitmanEtAl-2006}. Eventually most of the volatile material contained in a comet nucleus sublimates away, and the comet then becomes an inert lump of rock that resembles an asteroid, similar to a considerable fraction of near-Earth objects \citep{BinzelEtAl-2004}. There are three fundamental differences between comets around the sun and around a WD. First, the typical perihelion of the former is of the order of AU, whereas the tidal disruption radius of a typical WD is about 300 times less. Since the irradiation is proportional to the square of the orbital distance, disrupted comets that approach perihelion around a WD receive $\sim$10$^5$ times the intrinsic luminosity of the star. The second difference is the intrinsic luminosity. Compared to the luminosity of the sun, which is relatively constant during the main sequence and only changes by an order of unity, the luminosity of a WD is highly variable and decreases with the WD's cooling age. According to \cite{Veras-2016}, it varies between 10$^3$L$_\odot$ immediately after the AGB phase and decreases to 10$^{-5}$L$_\odot$ when the WD cools below 6000-8000 K, which corresponds to a cooling age of a few Gyr. The conclusion is that at the limit of L=10$^{-5}$L$_\odot$, tidally disrupted cometesimals around WDs should experience the same conditions as solar system comets do. Naturally, at lower cooling ages, the amount of irradiation could be up to 8 orders of magnitude higher, yielding more extreme conditions. Only for old WDs, the conditions could be more favorable for comets to retain their water. The third difference is that in the solar system, the sublimated gas from comets is dispersed by radiation pressure, whereas for virtually any WD (with $L<0.1L_\odot$) the radiation forces are too feeble compared to the gravitational forces and are thus unable to push away the gas \citep{BonsorWyatt-2010}, which will inevitably be accreted onto the WD.

Therefore, in the view presented by \cite{VerasEtAl-2014b}, a tidally disrupted minor planet will be rapidly disassociated into much smaller particles even compared to a typical solar system comet. The disassociated planetesimals would be scattered along the original trajectory of the disrupted minor body, which would be initially eccentric. At each periastron pass, the water in these fragments would rapidly sublimate. The water sublimataion rate would be largely dictated by the rate of the dissociation and the size of the fragments. According to the above arguments, in WDs with a cooling age of up to several Gyr, the conditions that such fragments encounter will be at the very least comparable to those of typical solar system comets, and full water loss would occur extremely fast, especially in young WDs or if the fragments are as small as those considered by \cite{VerasEtAl-2014b}, on the scale of hundreds of meters in diameter. We also note that comparable fragment sizes have been inferred for the tidally disrupted comet Shoemaker-Levy 9 \citep{Crawford-1997}. For fragments that small, the water might not survive more than several orbits. In fact, there is also the possibility that water sublimation and accretion begins even prior to a tidal disruption event. E.g., if a minor planet is perturbed gradually, and it initially has a periastron outside the tidal disruption radius, however sufficiently close to the star to enable water sublimation, then water begins to accrete onto the WD even before the disk begins to form. In that particular case we might expect relatively low rates of intermittent water accretion, corresponding to each periastron pass.

Finally, the actual accretion of the ensuing water gas onto the WD is believed to involve ionization of the gas which is then subject to magneto-rotational instability \citep{KingEtAl-2007}. The gas viscous dissipation timescale is expected to be (up to 6-7) orders of magnitude shorter than that resulting from PR drag \citep{MetzgerEtAl-2012,FarihiEtAl-2012,Farihi-2016}, which primarily governs the circularization time of the remaining silicate disk. In summary, this scenario predicts rapid sublimation of volatiles followed by rapid accretion. In some cases, the sublimation might even start prior to the start of disk formation via tidal disruption. Sublimation and accretion of refractory materials in large quantities occurs only after the formation and circularization of the disk, which, according to the above arguments may be several orders of magnitude slower. This scenario is also consistent with the observation that a considerable fraction of polluted He-dominated WDs exhibit average accretion rates that exceed that of polluted H-dominated WDs by one or two orders of magnitudes. The rate in H-dominated WDs is considered to be in agreement with the theoretical mass transfer limit via PR drag \citep{GirvenEtAl-2012}. Therefore the increased average accretion rate in He-dominated WDs is believed to involve intense and brief accretion episodes \citep{FarihiEtAl-2012}. So far several possible explanations have been suggested, all involving gaseous debris. The first possibility is related to the runway accretion model of \cite{Rafikov-2011}, which ends in an exponential burst of accretion of a large portion of the initial disk mass \citep{MetzgerEtAl-2012}. The other possibility is related to gas produced by collisions with an existing disk, either at the very early stages of its formation, immediately following the tidal disruption event \citep{FarihiEtAl-2012,BearSoker-2013}, or during later times as a result of the interaction of an impacting minor planet with the disk \citep{Jura-2008}. Here we propose that differential accretion, i.e., the early rapid accretion of sublimated volatiles, could be a simple and consistent explanation to this problem that hasn't been suggested before. Also, according to \cite{Farihi-2016}, the observational data favor accretion burst scenarios that allow for persistent (in this case silicate) debris.

\subsection{Predictions}\label{SS:predictions}
From Section \ref{SS:scarcity} we conclude that the most likely explanation for the relative scarcity of ongoing water accretion events is related to the early and rapid accretion of water during the initial stages of disk formation, or even prior to the beginning of disk formation by a tidal disruption event. Many of the high sensitivity observations of WDs involve polluted H-dominated atmospheres. We note that overall, H-dominated atmospheres are more common than their He-dominated counterparts \citep{AlthausEtAl-2010,Koester-2013}. Since their sinking time is typically on the scale of only days or weeks, the probability to observe water in their atmospheres would be approximately the ratio between the MRI gas accretion and PR drag time scales \ref{sss:DiffAccretion}, hence, very low (although according to \cite{Farihi-2016} the exact ratio may vary). Sequenced accretion does not explain the scarcity of water detections in He-dominated atmospheres, characterized by long sinking times, however it is consistent with the high average accretion rates observed in these WDs, implying brief episodes of rapid accretion beyond the limit of PR drag. 

For polluted He-dominated WDs, our best hypothesis for the scarcity of water detections is that the disrupted minor planets are intrinsically dry, or relatively dry. This could be explained if minor planets are depopulated from the inside out, since there is an observational bias for the detection of oxygen in younger, warmer WDs. Our results in Section \ref{S:Results} do indicate that minor planets at inner orbits become dry as a result of the RGB and AGB stellar evolution, or at least relatively dry. This result is robust for various combinations of model parameters, however large minor planets have a bigger chance to retain a significant fraction of the initial water, and therefore the scarcity of water detections in He-dominated WDs may also be related to the size distribution of these minor planets.

Overall we predict that future detections of accreted water-bearing minor planets are far more likely to involve He-dominated atmospheres than H-dominated atmospheres. In case water is inferred to contribute to the pollution of H-dominated atmospheres, we predict that it would most likely involve accretion of hydrated silicates from a long-lived disk, and therefore limit the water mass fraction to no more than 10-20\%. Indeed GD 61 and SDSSJ1242, which are the only two WDs thus far discovered to accrete 26\% and 38\% water mass fraction respectively, have He-dominated atmospheres.

\subsection{Planet habitability}\label{SSS:habitability}
Since this paper contributes to understanding water retention in planetary systems, it might also bear relevance for planet habitability around WDs. A terrestrial planet orbiting at 0.01 AU around a non-magnetic, relatively cool WD could be potentially habitable for 3-8 Gyr, providing ample time for life to develop \citep{Agol-2011,FossatiEtAl-2012}. At such a distance from the star however, a planet would not survive the post main sequence stellar evolution and therefore it must be perturbed into that orbit after the formation of the WD. According to \cite{NordhausSpiegel-2013} this scenario entails a circularization of the planet's orbit through tidal interaction, which would dissipate a huge amount of heat that would render it uninhabitable. \cite{BarnesHeller-2013} further argue against habitability due to tidal heating, if the planet has an eccentricity as small as $\sim 10^{-5}$. That puts constrains on the planetary configuration, since any additional planets could raise the eccentricity above this value. Moreover, they argue that UV emission from a young WD would rapidly cause water loss, thereby young WDs (up to a cooling age of a few Gyr) would also be incompatible with sustaining life. 

If these arguments are correct, they imply that the most likely path to habitability is to perturb and capture a single terrestrial planet in a close and circular orbit around a relatively old, non-magnetic WD. This planet could later accrete water-rich minor planets and become suitable for life. Our results are encouraging in that they certainly support an abundance of water-rich minor planets of various sizes. Moreover, the scarcity of water detections in polluted He-dominated WDs might suggest (see previous Section \ref{SS:predictions}) that minor planets are depopulated from the inside out, which increases the chance for perturbed water-rich minor planets in older WDs. The probability of a tidal crossing planetesimal for direct impact onto the planet should be on the same order of magnitude as the probability for direct impact onto the WD, since the planet's orbit is close to the tidal disruption radius, the planet has a similar size and the WD provides the gravitational focusing. Thus, the cross section for direct collision in a random encounter is linear in radius \citep{DebesSigurdsson-2002}, and the probability is only $\sim 0.01$. The water-bearing atmosphere of SDSSJ124231 has been inferred to accrete a body equivalent in size to Ceres, amounting to $\sim 10^{24}$ g \citep{RaddiEtAl-2015}. This is similar to the amount of water mass in the Earth's oceans. Therefore it demonstrates that a considerable amount of water can be accreted in relatively short time periods. Other WDs, that are more similar to GD 61, have been inferred to accrete a much lower mass, approximately $\sim 10^{21}$ g (measured masses are lower limits of the total mass). This means that the necessary water for habitability would depend on the exact flux of perturbed plnetesimals, a subject requiring further study.

\section{Conclusions}\label{S:Conclusions}
Our main goal in this paper has been to investigate the retention of water in planetary bodies during the RGB and AGB stellar evolution phases, using a state of the art thermo-physical model. This model is capable of addressing several shortcomings in similar studies, previously made. Primarily, our model has the capability to better characterize the impact of the main sequence evolution, on the RGB and AGB evolutions, by considering not merely the size and the orbital distance of a minor planet, but also its initial composition and its formation time. On top of this, this model is not merely a thermal evolution model, but instead a complete thermo-physical model. It fully and consistently considers the transport of mass and energy inside the body, phase transitions of various species, including the hydration and dehydration of rocks through interaction with liquid water, and a more realistic treatment for the density and the internal porous structure, based on our most recent understanding from Solar system research. 

The results presented here reaffirm the notion that water in minor planets could certainly survive the intense luminosity of a transforming star. We find that previous studies either under-estimate or over-estimate the amount of surviving water. Our main results are that even small, comet sized minor planets, may retain much, if not all of their initial water content, in exo-solar Kuiper belts. At orbital distances as close as 5 AU, some minor planets may retain up to 50\% of their initial water content, depending on several parameters. This water may either be retained as ice, shielded in the interior, or in the form of hydrated minerals.

Finally, we discuss the apparent scarcity of detections of individual accretion events of water-bearing minor planets. We suggest several possible explanations, and conclude that the most likely explanation is that perturbed minor planets that have an icy interior, suffer the same fate as Solar system comets, and rapidly lose their water during the early stages of disk formation, or even prior to a tidal disruption event. We therefore predict that water detections exceeding about 20\% in mass fraction will be extremely rare in polluted H-dominated WDs, since they do not record traces of past water accretion in their atmospheres due to their very short sinking time. Our model shows that minor planets larger than about 50 km in radius are nevertheless capable of undergoing an internal chemical evolution that incorporates some water onto the rocks, and therefore it may still be present in the silicate portion of the disk. Since in reality most perturbed minor planets should be smaller than 50 km, and the silicate disk could be composed of multiple disrupted bodies, water pollution from the silicate disk would most likely amount to much less than 20\% in mass fraction. In polluted He-dominated WDs a couple of water-rich atmospheres have already been detected, however most He-dominated WDs are not observed to be water-rich. This could be explained via the observational bias that all of these WDs are relatively young. Young WDs might be polluted by intrinsically dry or relatively dry minor planets due to their depopulation sequence, although more research is required in order to determine the most likely perturbation mechanisms.  

Overall, we predict that most if not all water detections in the near future would involve He-dominated atmospheres. We also discuss how water retention in minor planets may enable the habitability of a single terrestrial planet around old, non-magnetic WDs. This paper emphasizes the need for additional observations of polluted WDs with high spectral sensitivity, which would permit better characterization and statistics in the future, as well as additional theoretical studies regarding planetary configurations in exo-solar planetary systems and the perturbation, size distribution and accretion of minor planets.

\section{Acknowledgment}\label{S:Acknowledgment}
We would like to thank the anonymous reviewer for valuable comments and suggestions to improve the quality of the paper. We also wish to thank Dimitri Veras for reviewing the manuscript and providing helpful comments and suggestions, as well as Jay Farihi and his research group. UM and HBP acknowledge support from the Marie Curie FP7 career integration grant "GRAND", the Research Cooperation Lower Saxony-Israel Niedersachsisches Vorab fund, the Minerva center for life under extreme planetary conditions and the ISF I-CORE grant 1829/12.

\newpage


\bibliographystyle{apj} 

\end{document}